\documentclass[letterpaper, 10 pt, conference]{ieeeconf}  
\usepackage{xcolor}

\usepackage{amsmath,amssymb,amsfonts}
\usepackage{tabularx,booktabs,graphicx,multirow}
\usepackage[caption=false]{subfig}

\usepackage[maxfloats=256]{morefloats}
\newcommand{\COtwo}{\mathrm{CO}_2}
\newcommand{\msq}{\mathrm{m}^2}
\newcommand{\degrees}{^\circ\mathrm{C}}
\newcommand{\Tindoor}{\mathrm{T}_\mathrm{indoor}}
\newcommand{\Toutside}{\mathrm{T}_\mathrm{outside}}
\newcommand{\RH}{\mathrm{RH}}
\newtheorem{remark}{Remark}

\makeatletter
\let\NAT@parse\undefined
\makeatother
\usepackage[colorlinks=true,linkcolor=blue!80!black,citecolor=blue!80!black,urlcolor=blue!80!black,breaklinks]{hyperref}
\IEEEoverridecommandlockouts

\title{\LARGE \bf
What influences occupants' behavior in residential buildings:  An experimental study on window operation in the KTH Live-In Lab
}

\author{
Mahsa Farjadnia$^{1,\star}$, Angela Fontan$^2$, Alessio Russo$^2$, Karl Henrik Johansson$^2$, Marco Molinari$^{1}$%
\thanks{This work was supported by the Swedish Energy Authority and IQ Samhällsbyggnad, under the E2B2 program, grant agreement n. 47859-1 (Cost-and Energy-Efficient Control Systems for Buildings), by SSF (Swedish Foundation for Strategic Research), under grant agreement n. RIT17-0046 (CLAS--Cybersäkra lärande reglersystem), by the Swedish Research Council Distinguished Professor Grant 2017-01078, and by the Knut and Alice Wallenberg Foundation Wallenberg Scholar Grant. The KTH Live-In Lab was made possible by a donation from the Einar Mattsson-Group, whose support is kindly acknowledged. Donations from Akademiska Hus and Schneider Electric are also kindly acknowledged.}%
\thanks{$^1$%
M. Farjadnia and M. Molinari are with the Division of Applied Thermodynamics and Refrigeration, KTH Royal Institute of Technology, 100 44 Stockholm, Sweden, E-mail: \{mahsafa,marcomo\}@kth.se.}%
\thanks{$^2$%
A. Fontan, A. Russo, and K. H. Johansson are with the Division of Decision and Control Systems, KTH Royal Institute of Technology, 100 44 Stockholm, Sweden, E-mail: \{angfon,alessior,kallej\}@kth.se.}%
\thanks{The authors are  also affiliated with Digital Futures.}
\thanks{$^\star$Corresponding author.}%
}

\begin{document}
\bstctlcite{IEEEexample:BSTcontrol}

\maketitle
\thispagestyle{empty}\pagestyle{empty}

\begin{abstract}
Window-opening and window-closing behaviors play an important role in indoor environmental conditions and therefore have an impact on building energy efficiency. On the other hand, the same environmental conditions drive occupants to interact with windows. Understanding this mutual relationship of interaction between occupants and the residential building is thus crucial to improve energy efficiency without disregarding occupants' comfort. This paper investigates the influence of physical environmental variables (i.e., indoor and outside climate parameters) and categorical variables (i.e., time of the day) on occupants' behavior patterns related to window operation, utilizing a multivariate logistic regression analysis.
The data considered in this study are collected during winter months, when the effect on the energy consumption of the window operation is the highest, at a Swedish residential building, the KTH Live-In Lab, accommodating four occupants in separate studio apartments. Although all the occupants seem to share a sensitivity to some common factors, such as air quality and time of the day, we can also observe individual variability with respect to the most significant drivers influencing window operation behaviors.
\end{abstract}

\section{INTRODUCTION}
Energy demand in the residential and commercial buildings sector in the European Union accounts for approximately 40$\%$ of total end-use energy \cite{webEC} and 76$\%$ of this energy is used in comfort control in buildings using HVAC (heating, ventilation, and air conditioning) systems \cite{perez2008review}. The behavior of building occupants, specifically window-opening actions, plays a fundamental role in the amount of energy used in buildings \cite{fabi2012occupants}, particularly during the heating season. Previous research indicates that building energy consumption can be reduced if the control strategies of HVAC systems are optimized considering how individuals behave \cite{dai2020review}. Therefore, efforts have been devoted over the years to developing accurate modeling of occupant behavior \cite{barthelmes2017exploration,cali2016analysis,hong2016advances}. In this work, we argue that a better understanding of the drivers that motivate occupants to interact with windows in smart buildings is significant in order to improve the efficiency of recently developed control strategies, in particular,
 data-driven predictive control techniques \cite{khosravi2019data,di2022lessons,farjadnia2022robust,yang2020hvac}.
 
\subsection{Related Work}
Existing literature on window-opening behavior can be classified based on the applied theoretical methods, e.g., logistic regression \cite{haldi2009interactions,andersen2013window,shi2020effects}, Markov processes \cite{haldi2009interactions,schweiker2012verification}, deep learning methods \cite{markovic2018window}, and the type of building under study, residential \cite{rouleau2020probabilistic,jones2017stochastic} or office buildings \cite{markovic2018window,markovic2017comparison,li2015probability}. A detailed review is provided in \cite{dong2018modeling}, where various modeling techniques are categorized based on their complexity, data requirements, and level of implementation. Logistic regression was reported among the widely implemented and relatively simple models. 
A remarkable contribution in modeling window behavior can be found in \cite{haldi2009interactions}, where the authors applied logistic regression, a Markov process, and a combination of these approaches to model the relationship between the window behavior of occupants and environmental drivers in office buildings, using seven years of data. The authors concluded that indoor temperature was a significant driver in the window-opening action, and outdoor conditions best described the window-closing action.

Although most of the existing literature is focused on office buildings (for a review \cite{fabi2012occupants}), the interest in analyzing occupants' behavior in residential buildings is increasing. 
In \cite{jones2017stochastic} the authors assessed the window-opening action of occupants of a residential building in the UK using one-year data considering environmental drivers, time of the day, and season. The authors of \cite{shi2020effects} utilized a multilevel logistic regression model to evaluate the effect of household features, such as area and type of residents (i.e., smoker and age), in the window-opening behavior of occupants in 10 apartments. Features leading the occupants to interact with windows in 60 residential buildings are studied in \cite{cali2016analysis}. Time of the day, carbon dioxide concentration, and average outside temperature were reported as the most common drivers in opening and closing the windows.

Although some of the key challenges have been identified in previous studies, it is important to emphasize that there is still no consensus on what types of drivers (physical environmental, physiological, social, and psychological) lead to an interaction between occupants and buildings. Moreover, existing literature primarily focuses on analyzing drivers of actions in multi-occupancy residential buildings, which only indicates the general behavior of the occupants and does not reveal individual differences between each occupant \cite{jian2022individual}.

\subsection{Contributions}
This work aims at modeling occupants' behavior in the context of smart homes, by determining the most statistically significant drivers that can be used to identify case-specific behaviors. 
In particular, we focus on the interaction between occupants and windows during the winter months, when the effects of behavior patterns are most noticeable in energy consumption. To this end, we consider two  models distinguishing between window-opening and window-closing actions. By building separate models, we can capture the specific characteristics associated with each action more effectively. We investigate the influence that drivers such as indoor temperature, carbon dioxide ($\COtwo$) consumption, indoor relative humidity (RH), outside temperature, and time of the day have on the window-opening and window-closing behavior of the occupants at the KTH Live-In Lab.

The KTH Live-In Lab infrastructure is designed to offer a comprehensive infrastructure (to collect, store, and visualize data) and dedicated experimental setup with state-of-the-art monitoring capability for residential buildings \cite{Molinari2022LongTerm}, previously available mainly (if not only) in office spaces. The results obtained in this research then contribute to the growing literature focusing on household behavior and on the dynamic interaction between occupants and the residential building.
In addition, differently from the aforementioned studies (e.g., \cite{cali2016analysis}) where apartments typically have not only a larger floor surface but also more than one occupant per apartment, in the experimental setup of the KTH Live-In Lab the studios have (on average) a floor surface of 20$\msq$ and are leased to a single occupant. This particular configuration allows us to study not only influential drivers in \textit{residential} buildings, but also \textit{heterogeneity} among tenants. The motivation is that, as tenants do not have direct control over the air quality settings, insights gained in this study regarding the individual occupant characteristics and preferences will serve as a foundation 
for developing customized controller designs capable to capture individual needs and comfort levels.

To model the occupant's window operation behavior we employ a logistic regression technique, following a stepwise regression procedure similar to the one illustrated in \cite{cali2016analysis}. We consider the aforementioned explanatory variables, i.e., indoor and outside temperatures, RH, and $\COtwo$. In addition, we add a categorical variable to represent different day segments where different probabilities of interaction with windows (opening or closing) were observed. Our results show that the most common driver for tenants to open a window during the winter season is the day segment, followed by $\COtwo$. For window-closing behavior, the most common drivers are the day segment, and indoor and outside temperatures.
The classification models are evaluated using different performance indicators, such as the ROC (receiver operating characteristic) curves and the precision-recall curves.

During data analysis and processing, we recognized typical challenges of collecting and working with real data from a testbed, which might have influenced the predicting capabilities of the computed models. The two most common were missing data (lost in the recording process), which means that the available number of measurements in the same apartment differed between each sensor, and imbalanced data, which means that the number of times tenants interacted with windows was low (w.r.t. summer months).

Apart from these practical issues, difficulties to model the tenants' actions may also be explained by unexpected behavioral patterns that we discovered after the experimental study ended, through a discussion with the participants. For example, it was revealed that there were instances where a tenant would keep a window open, but at the same time would leave an electrical radiator on. 
This is consistent with representing the KTH Live-In Lab as a complex cyber-physical-human system \cite{CPHS2022,Lamnabhi-Lagarrigue2017Systems}. It is not the scope of this paper, but future directions will consider further challenges that arise from the interplay between the human component (i.e., the tenants) and the cyber-physical systems (i.e., the building). For instance, we plan to investigate the role and effect of social influence in environmentally significant behaviors \cite{Fontan2023SocialInteractions}.

\subsection{Outline}
The paper is organized as follows: Section~\ref{sec:infrastructure}  introduces the experimental setup and gives a brief overview of the
KTH Live-In Lab building, data infrastructure, and tenants; Section~\ref{sec:modeling_behavior} describes the modeling and data preprocessing procedure; Section~\ref{sec:results} presents the two obtained models for window-closing and window-opening behavior; finally, Section~\ref{sec:conclusion} offers conclusive remarks.

\section{THE KTH LIVE-IN LAB}
\label{sec:infrastructure}
\begin{figure}[b]\centering
\includegraphics[width=.48\textwidth]{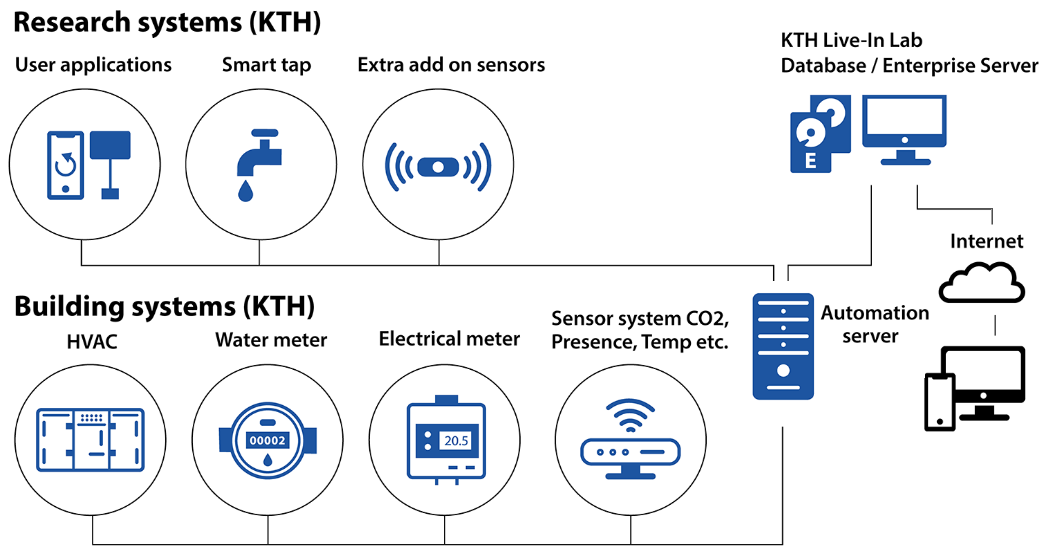}
\caption{Simplified illustration of the KTH Live-In Lab data infrastructure. Image source: \href{https://www.liveinlab.kth.se/en/infrastruktur/datapool/data-infrastructure}{liveinlab.kth.se/en/infrastruktur/datapool/data-infrastructure}.}
\label{fig:data-infrastructure}
\end{figure}

The KTH Live-In Lab includes several building testbeds that range from students' accommodations to lecture buildings \cite{molinari2023using}.  Data collected from the sensors installed in the testbeds is stored and shared through the KTH Live-In Lab datapool (Fig.~\ref{fig:data-infrastructure}). The KTH Live-In Lab is an example of cyber-physical-human system, featuring a redesignable testbed layout (Section~\ref{sec:layouts}), an extended sensor network (Section~\ref{sec:sensors}), and an advanced interaction capability with testbed occupants (Section~\ref{sec:tenants}).

\subsection{Building Infrastructure}\label{sec:layouts}

\begin{figure*}[t]\centering
\subfloat[]{\includegraphics[height=4.1cm,trim={0 0 0 0.5cm},clip]{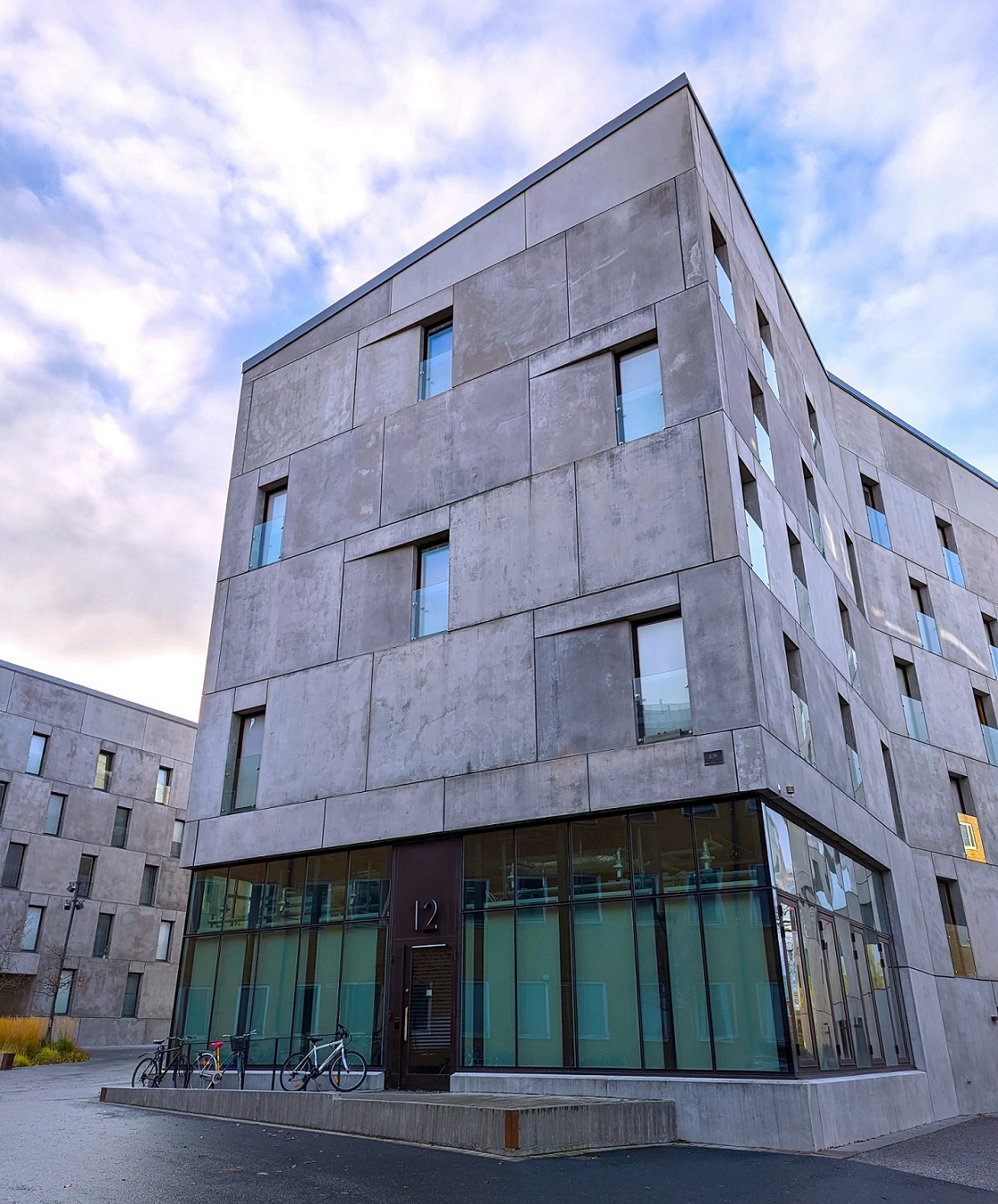}\label{fig:Outside}}\;
\subfloat[]{\includegraphics[height=4.1cm]{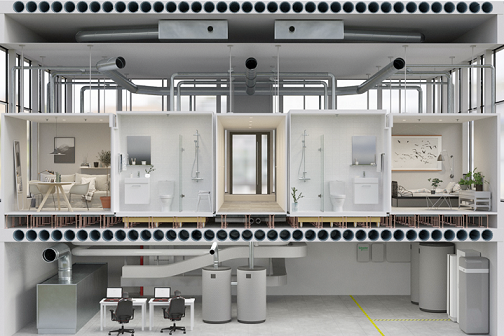}\label{fig:sectional-view}}\;
\;
\subfloat[]{\includegraphics[height=4.2cm]{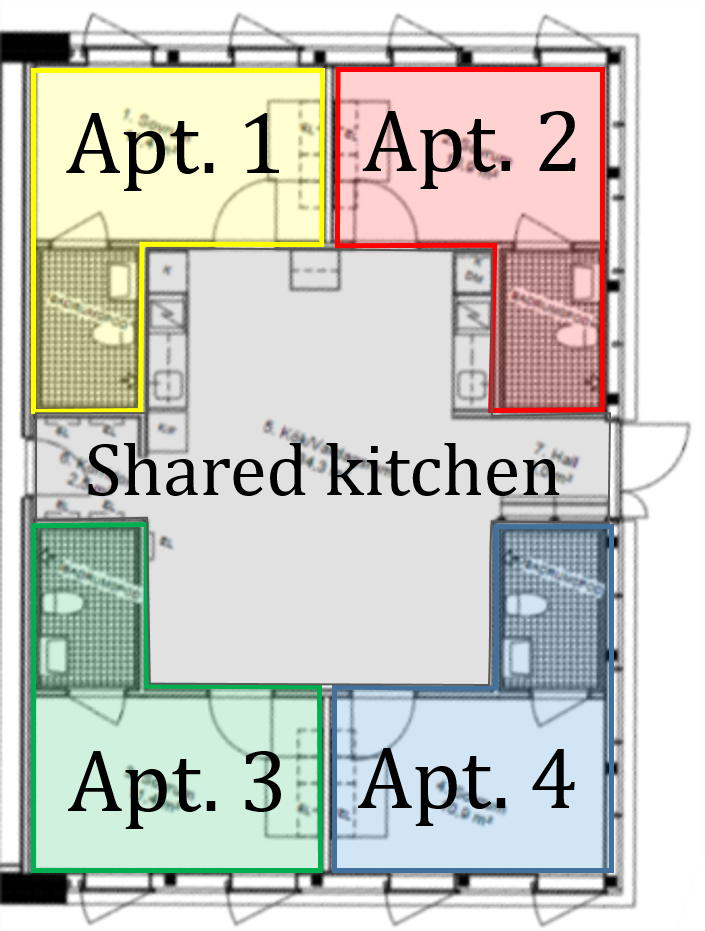}\label{fig:Testbed2}}
\,\subfloat[]{\includegraphics[height=4.2cm]{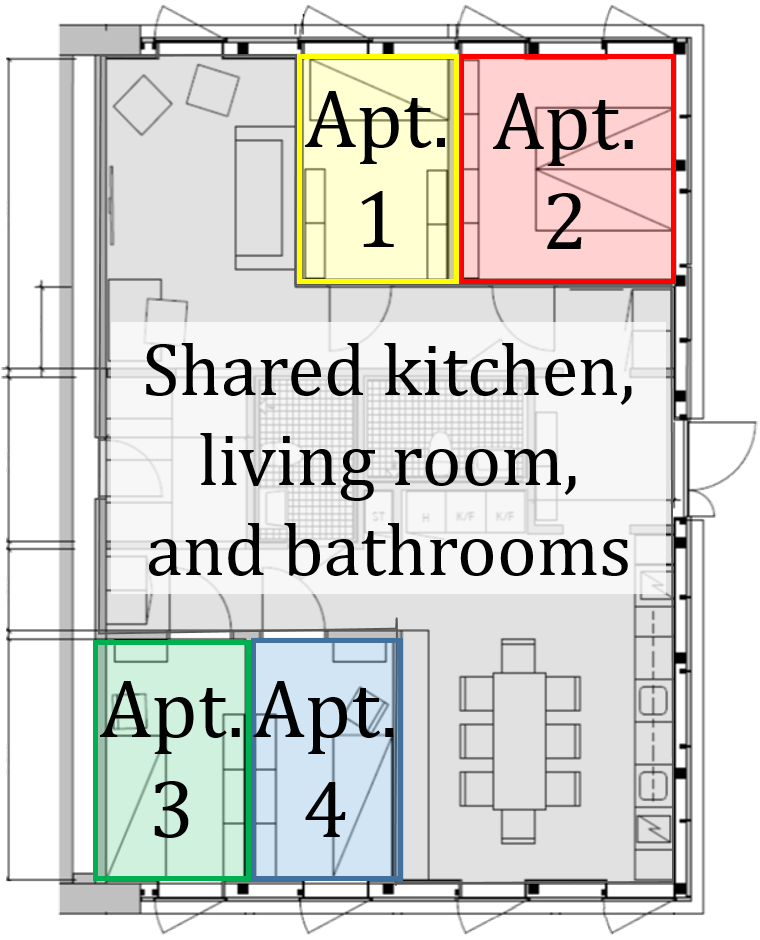}\label{fig:Testbed3}}
\\
\subfloat[]{\includegraphics[height=3.8cm]{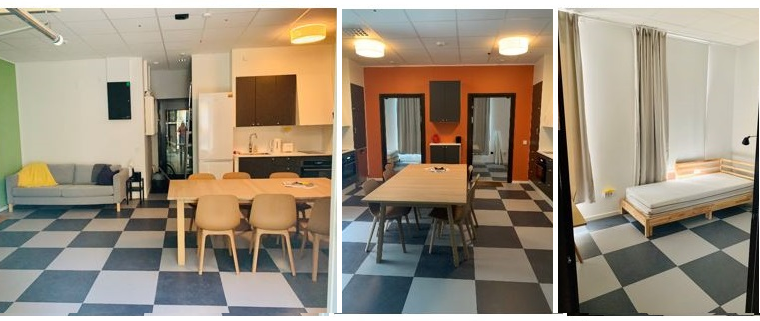}\label{fig:Testbed2-Inside}}
\;\,
\subfloat[]{\includegraphics[height=4cm]{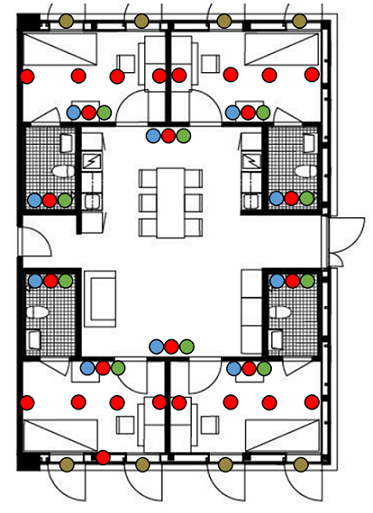}\!\!
\includegraphics[height=3.6cm]{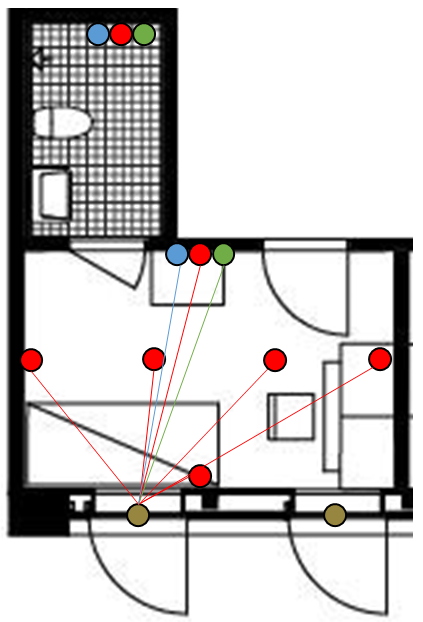}\label{fig:sensors-testbed2}}
\;\subfloat[]{\includegraphics[height=4cm]{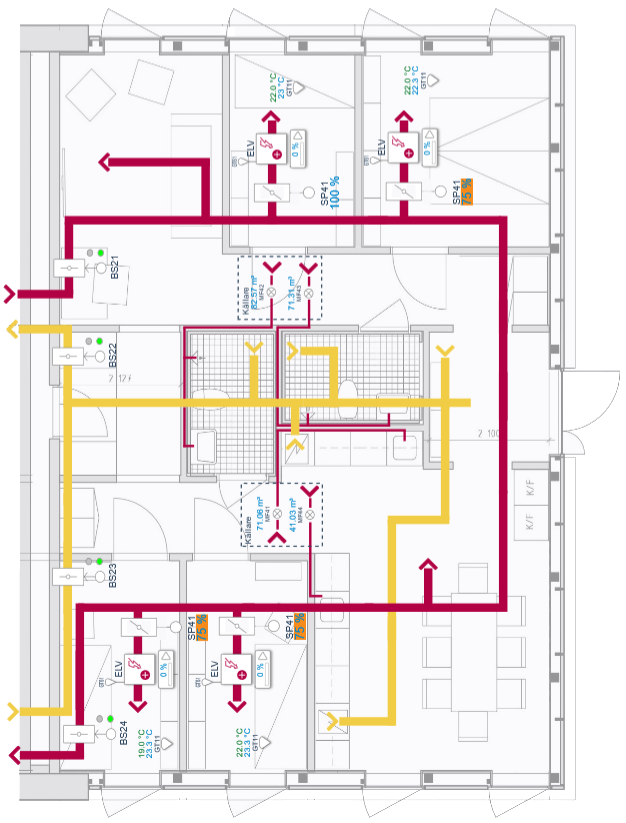}\label{fig:Testbed3-ventilation}}
\caption{(a): Outdoor view of the KTH Live-In Lab. (b): Sectional view of the KTH Live-In Lab. (c): The layout of the KTH Live-In Lab applied from August 2020 up to June 2021. (d): The current layout of the KTH Live-In Lab applied since August 2021. (e): Inside view (common living space, kitchen, bedroom) of the KTH Live-In Lab before the summer of 2021. 
(f) Sensors network in the whole KTH Live-In Lab with the layout as in (c) (left) and in a single apartment (right). The depicted sensors are magnetic sensors used to detect window-opening (light brown), temperature sensors (red), RH sensors (blue), and $\COtwo$ sensors (green). Light, motion, and VOC sensors are not depicted.  (g): Ventilation system of the current layout of KTH Live-In Lab. Red and yellow colors indicate supply and extraction air ducts, respectively.
Images source: \href{https://www.liveinlab.kth.se/en/infrastruktur/testbed-infrastructure}{liveinlab.kth.se/en/infrastruktur/testbed-infrastructure}.}
\label{fig:TestbedEM}
\end{figure*}
The overall floor area of the KTH Live-In Lab is 300$\mathrm{m}^2$, see Figs.~\ref{fig:Outside} and~\ref{fig:sectional-view} for an outdoor view and a sectional view of the KTH Live-In Lab, respectively, and its layout can be redesigned (Figs. \ref{fig:Testbed2} and \ref{fig:Testbed3}). 
The layout of the KTH Live-In Lab before the summer of 2021 comprised four independent apartments, each consisting of a living room/bedroom and a bathroom, and a shared kitchen (Figs.~\ref{fig:Testbed2} and~\ref{fig:Testbed2-Inside}). In the current design, the concept of shared spaces was implemented by incorporating shared living room and bathrooms in addition to the shared kitchen, while still keeping private bedrooms (Fig.~\ref{fig:Testbed3}).  

It is worth noting that the heating is provided through a ventilation system which uses an air-handling unit, with dedicated distribution outlets for each apartment (Fig.~\ref{fig:Testbed3-ventilation}). This setup allows building managers to customize control over heating and air quality settings, while tenants have limited control of the system, constrained to the heating setpoints.

\subsection{Data Infrastructure}\label{sec:sensors}
The KTH Live-In Lab is equipped with advanced sensing technologies used to monitor indoor environmental parameters, such as indoor temperature, RH, $\COtwo$, and volatile organic compounds. 
Additional sensors include contact sensors deployed to detect, for instance, windows and doors status (open or closed), occupancy, and light sensors. These sensors are used to study solutions that optimize the use of energy for heating and ventilation, as well as maximize the use of daylight, and improve light comfort. All measurements are taken from sensors placed at apartment level (Fig.~\ref{fig:sensors-testbed2}). A detailed description can be found in \cite{Molinari2022LongTerm}.

\subsection{Tenants}\label{sec:tenants}
The tenants of the KTH Live-In Lab are undergraduate students attending KTH Royal Institute of Technology that reside at the KTH Live-In Lab full-time. Independently from the layout (Fig.~\ref{fig:TestbedEM}), each apartment is leased to a single tenant.

\section{MODELING OCCUPANTS' BEHAVIOR}
\label{sec:modeling_behavior}

\subsection{Development and evaluation of the models for opening-window and closing-window behavior}
Multivariate logistic regression was used to infer the probability of opening or closing a window based on a certain set of independent explanatory variables (e.g., indoor temperature or RH). In this paper, we use interchangeably the notation ``drivers'' and ``explanatory variables''. The probability $p$ of a window-opening or window-closing action is defined in terms of the logit function as:
\begin{align}\label{eqn:probability}
    \ln \left( {\frac{p}{{1 - p}}} \right) = \alpha  + {\beta _0}{x_0} + {\beta _1}{x_1} + \dots + {\beta _n}{x_n},
\end{align}
given $n$ explanatory variables denoted by $x_i$, $i = 1, \dots n$. In eq.~\eqref{eqn:probability}, $\alpha$ is the intercept, and $\beta_i$ is the coefficient of the explanatory variable $x_i$.

\begin{figure}[t]\centering
    \includegraphics[width=0.4\textwidth,page=2]{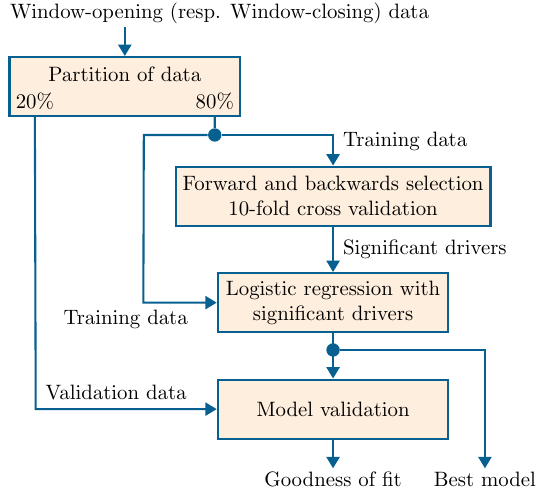}
    \caption{Diagram of the process for selecting the best fitting window-opening (resp. window-closing) model based on monitored window-opening (resp. window-closing) data, i.e., measurements collected when the window was closed (resp. open).}
    \label{fig:diagram}
\end{figure}

For the analysis we followed a procedure similar to the one illustrated in \cite{cali2016analysis}; however, differently from \cite{cali2016analysis}, we include an additional validation step  (Fig.~\ref{fig:diagram}). We report the details herein for completeness. Note that the following procedure was applied twice, first using window-opening data and then using window-closing data.

First, data was partitioned: 80$\%$ of the total data set for the selection of significant explanatory variables and model training, and 20$\%$ for model validation.

Then, the training data was partitioned into 10 sub-samples for cross-validation: 9 sub-samples to be used for model training through forward and backward selection, while 1 sub-sample to be used for model validation. The procedure was repeated 10 times to utilize each sub-sample once as a test sub-sample.
The Akaike Information Criterion (AIC) was used for selecting the explanatory variables in the regression model, to take into account the risk of overfitting and the risk of underfitting \cite{cali2016analysis}. In practice, the selection of the most suitable model given $n$ explanatory variables was executed by the \verb|stepwiseglm| function in \textsc{MATLAB} software, which operates the following steps. 
First, each individual variable is fitted by a regression model and, after calculating the AIC for each fit, the single-variable model with the lowest AIC is selected. Then, the remaining variables are fitted by a bivariate regression model, and the one with the lowest AIC is chosen and compared to the selected single-variable model. If the single-variable model has a lower AIC, it is chosen as the final model. Otherwise, three-variable models are considered. While a forward-selection procedure continues to check if adding variables may lead to a model with a lower AIC, in parallel a backward-selection procedure checks if a lower AIC can be obtained by removing variables from the models. The outcome of the forward and backward selection is a model with the lowest AIC.
In our case, this method was used to determine the most significant explanatory variables for window-opening or window-closing actions.

Once the significant drivers were singled out, all training data were used to fit a logistic regression model with determined explanatory variables. 

Finally, the validation data was used to evaluate the goodness of fit of the estimated model, defined in terms of area under the curve (AUC) of the ROC curve and AUC of the precision-recall curve. The value of AUC is between 0 and 1, with 1 indicating perfect classification.  
\subsection{Collection and Preprocessing of sensors data}\label{sec:data_processing}
\begin{figure}[t]\centering
\subfloat[]{\label{fig:box-plot}\includegraphics[width=0.48\textwidth]{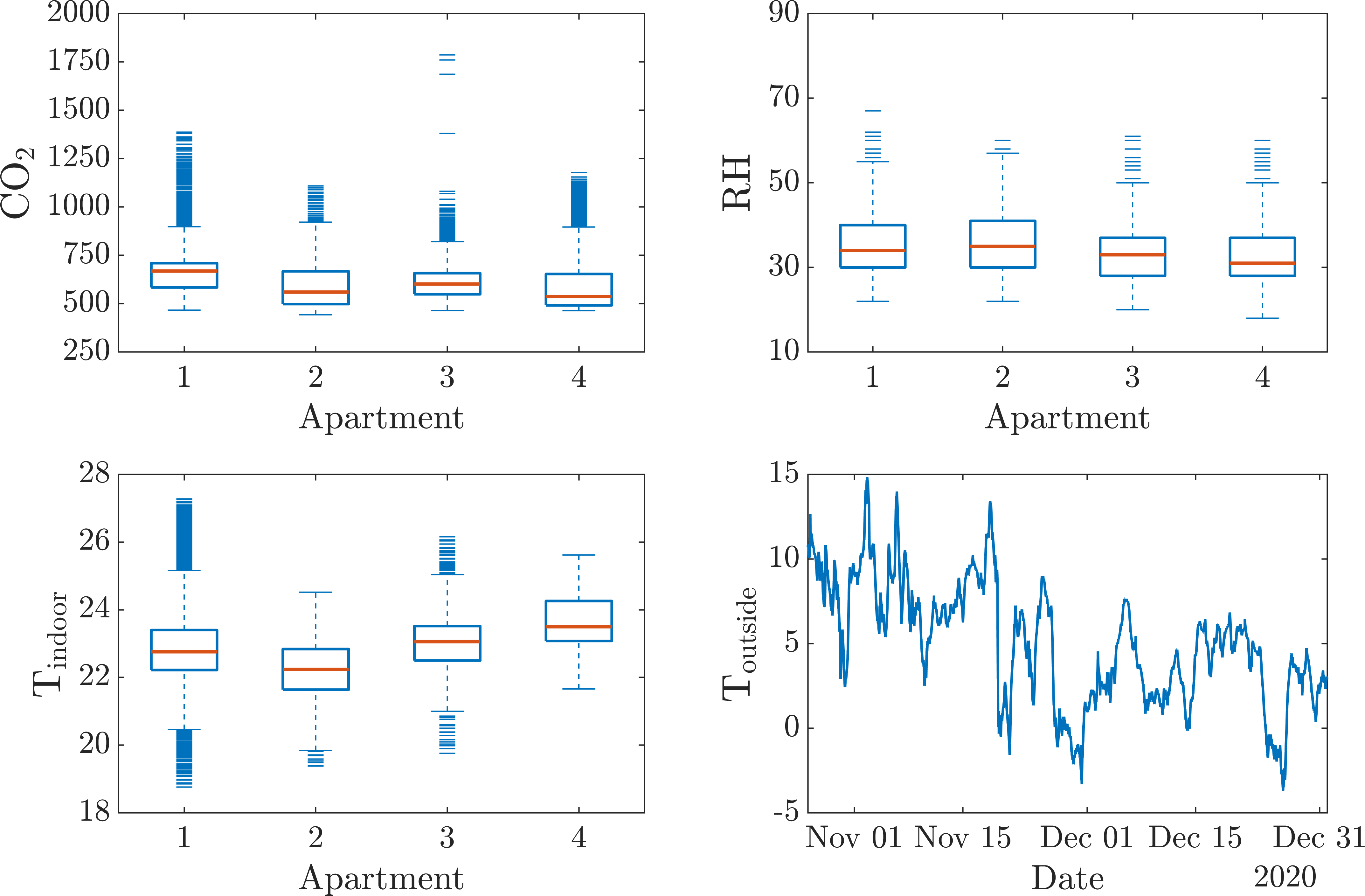}}\\
\subfloat[]{\includegraphics[width=0.47\textwidth]{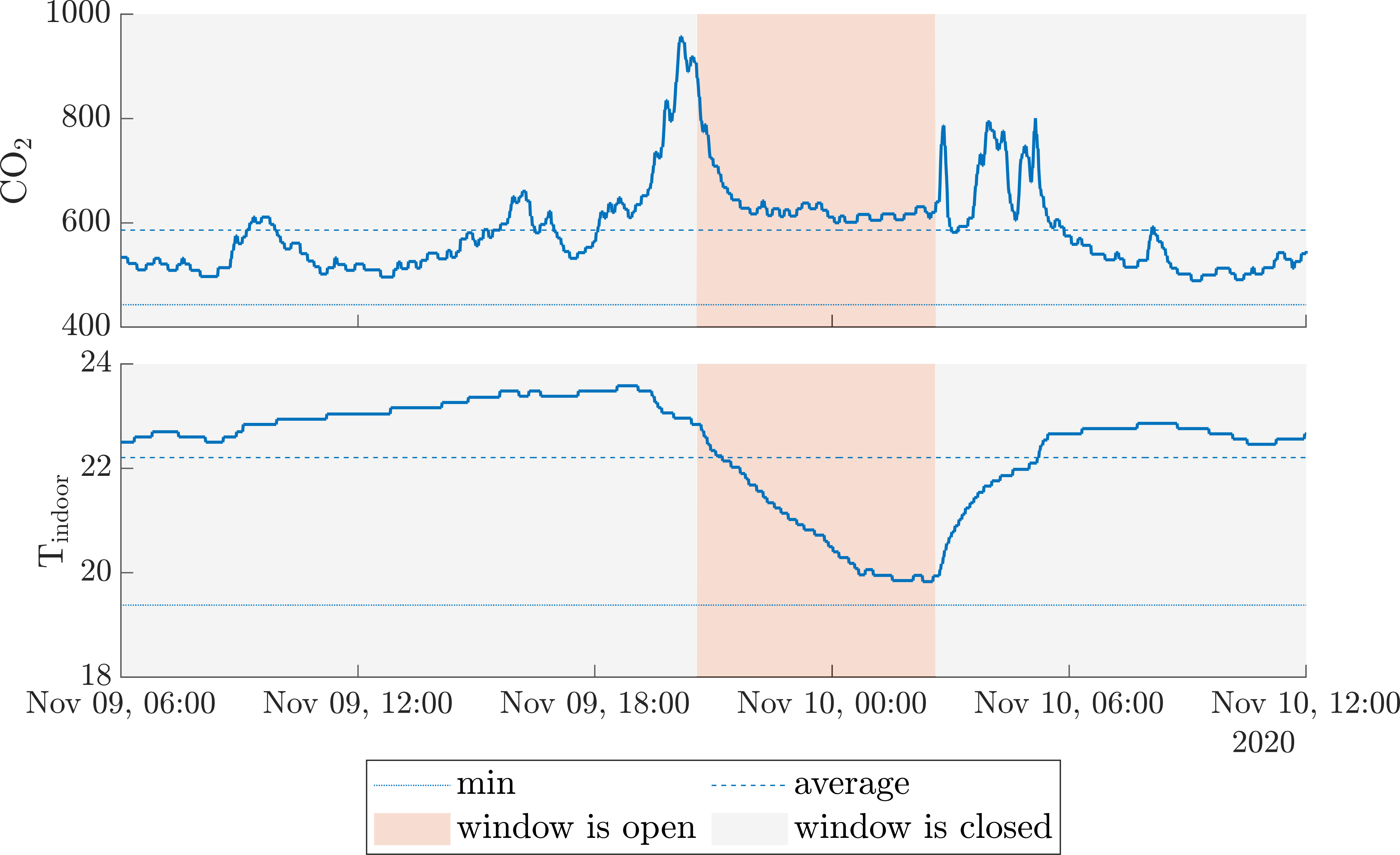}}
\caption{(a): Monitored data for the KTH Live-In Lab apartments during the period Nov-Dec 2020: $\COtwo$, RH, indoor temperature, and outside temperature. The box is plotted from the 1st to the 3rd quartile, the red line in the boxes represents the median, and the blue horizontal lines the outliers. The outside temperature can be seen in the bottom right panel. (b): Zoomed-in detail (Nov 9-10, 2020, apartment 2) to compare $\COtwo$ (top panel) and indoor temperature $\Tindoor$ (bottom panel) levels with opening- (red area) and closing- (gray area) window behavior.}
\label{fig:monitoredData}
\end{figure}

\begin{figure}[t]\centering
    \includegraphics[width=0.48\textwidth]{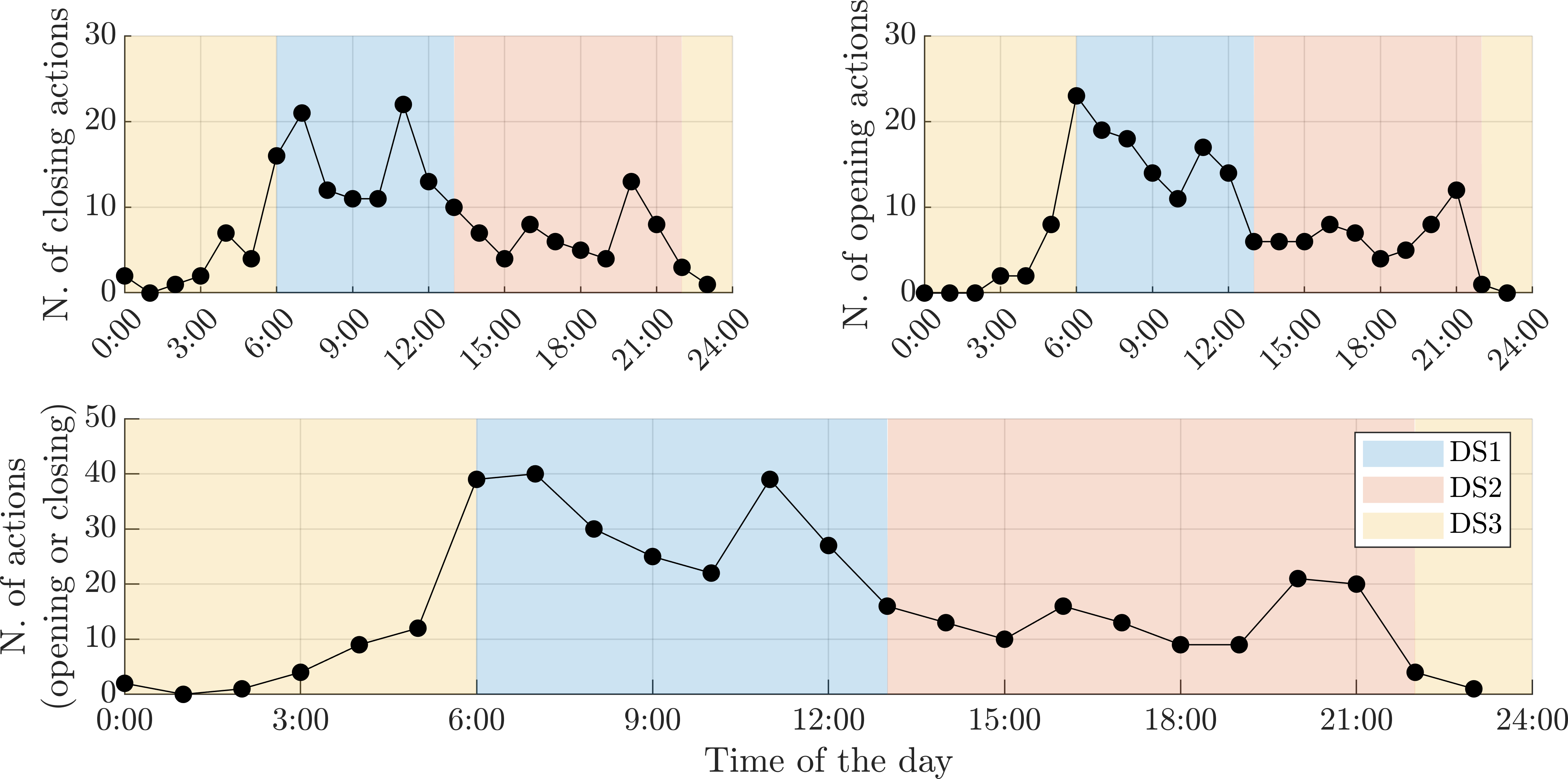}
    \caption{Observed total N. of times the windows were closed (top left panel), opened (top right panel), and opened or closed (bottom panel) in all apartments of the KTH Live-In Lab in Nov-Dec 2020, with respect to the time of the day. The three day segments (DS) are defined based on the probability of occupants interacting with a window (opening or closing action): DS1 $=$ [6:00-13:00), DS2 $=$ [13:00-22:00), and DS3 $=$ [22:00-6:00) (Section~\ref{sec:data_processing}). The probability of occupants interacting with a window is the highest during DS1, and the lowest during DS3.}
    \label{fig:windownum}
\end{figure}

The data used in this study was collected from October to December 2020, corresponding to the layout of the KTH Live-In Lab applied before summer 2021 (Fig. \ref{fig:Testbed2}). An illustration of the measured data is presented in Fig.~\ref{fig:monitoredData}. The sensors used in this experimental study are standard commercial sensors typically employed in building applications. These sensors have accuracy that is common for such purposes \cite{cali2016analysis}, e.g., the accuracy of the indoor and outdoor temperature sensors is within $\pm 0.5\degrees$ and $\pm 0.6\degrees$, respectively. The data was preprocessed before utilization due to missing values, incorrect measurements, and unsynchronized data resulting from differences in the sampling rate of different sensors. After preprocessing the data, the following variables with a sample time of one minute were considered:
\begin{itemize}
\item Window status (open/closed) magnetic sensors [-];
\item Indoor temperature ($\Tindoor$) $[\degrees]$;
\item Carbon dioxide concentration ($\COtwo$) [ppm]; log($\COtwo$) was used to have a better distribution for logistic regression analysis;
\item Relative humidity (RH) [\%];
\item Outside temperature ($\Toutside$) $[\degrees]$.
\end{itemize}
Based on observations of the window-opening activity in different apartments (Fig.~\ref{fig:windownum}), we decided to consider a categorical variable, denoted by day segment, which groups specific time periods throughout the day where difference in occupants behavior was most noticeable:
\begin{itemize}
    \item Day segment 1 (DS1): between 06:00 and 13:00;
    \item Day segment 2 (DS2): between 13:00 and 22:00;
    \item Day segment 3 (DS3): between 22:00 and 06:00.
\end{itemize}
The daily average outside temperature was used to avoid any correlation with the day segment variable.

\begin{figure}\centering
    \includegraphics[width=0.475\textwidth]{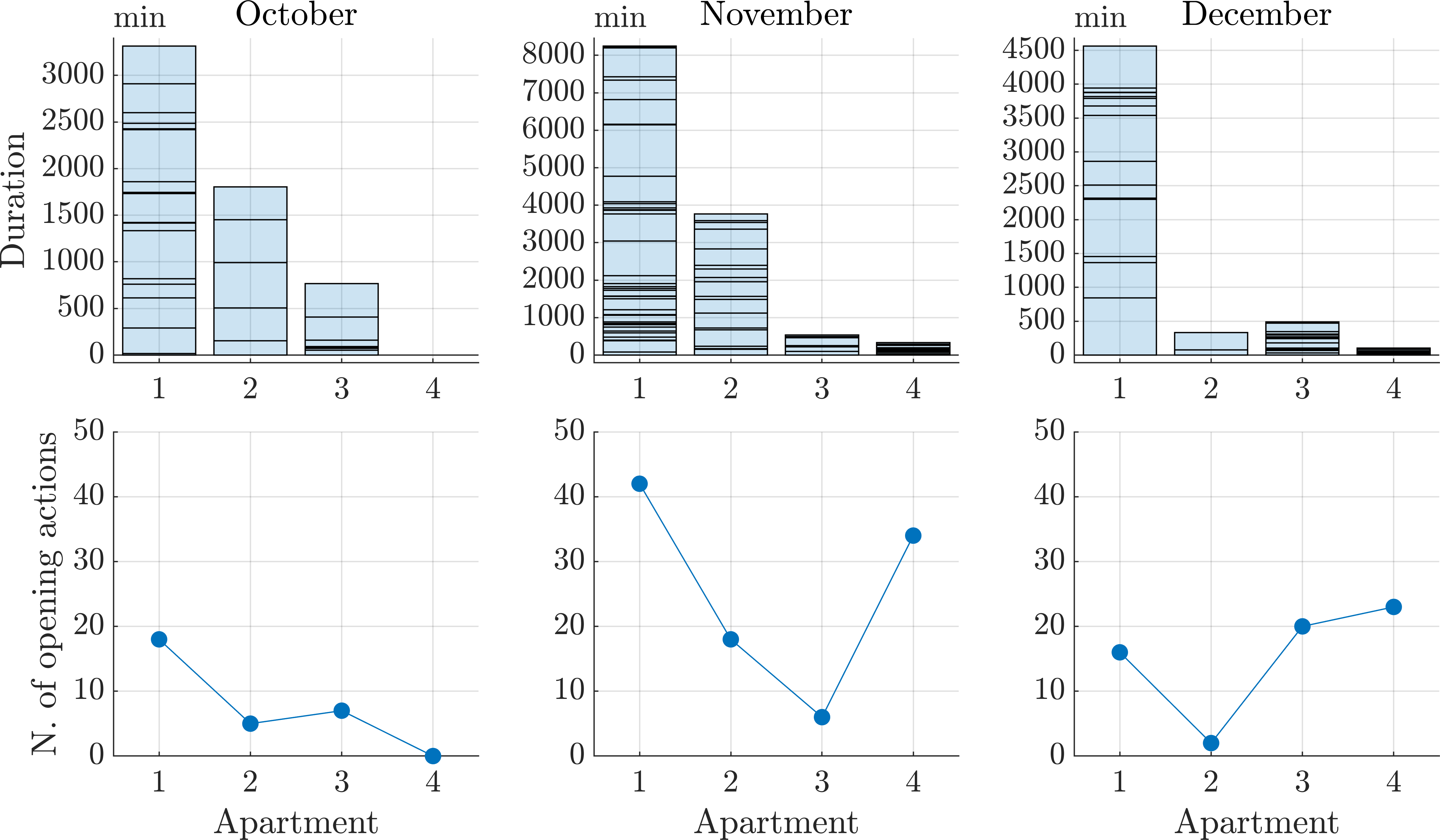}
    \caption{Observed total duration of open windows in number of minutes (top panels) and total N. of times the windows were open (bottom panels) in all apartments of the KTH Live-In Lab in Oct-Dec 2020, with respect to the month. In the top panels, horizontal black lines in each bar highlight the single durations of open window for each apartment across one month.}
    \label{fig:opening_duration}
\end{figure}

\section{EXPERIMENTAL RESULTS}\label{sec:results}
Following the procedure explained in Section \ref{sec:modeling_behavior}, we obtained a window-opening and a window-closing model for each apartment of the KTH Live-In Lab. For ease of exposition, we first illustrate the window-opening and window-closing models for apartment 2 (Section~\ref{sec:results-APT2}), and then we present the main results of the logistic regression analysis for all apartments (Section~\ref{sec:mainresults}). 

\subsection{Illustrative example of window-opening and window-closing models}\label{sec:results-APT2}
The models for window-opening and window-closing behavior for apartment 2 are reported in eq.~\eqref{eq:open-win} and eq.~\eqref{eq:close-win}, where the subscripts $o$ and $c$ stand, respectively, for opening and closing window action.
\begin{align}
\ln \left( \frac{p_{o}}{{1 - p_{o}}} \right) 
    &= \alpha_{o,\mathrm{DS}}  + \beta _{o,\Tindoor} x_{o,\Tindoor} \nonumber\\ &+ \beta _{o,\log(\COtwo)}x_{o,\log(\COtwo)}
      + \beta_{o,\RH} x_{o,\RH} \nonumber\\
        &+\beta_{o,\Toutside}x_{o,\Toutside},\label{eq:open-win}
      \\
\ln \left( \frac{p_{c}}{{1 - p_{c}}} \right) 
    &= \alpha_{c,\mathrm{DS}}  + \beta _{c,\Tindoor}x_{c,\Tindoor} \nonumber\\ &
      + \beta_{c,\RH} x_{c,\RH} 
      +\beta_{c,\Toutside} x_{c,\Toutside}.\label{eq:close-win}
\end{align}
The explanatory variables (Section \ref{sec:data_processing}) are denoted by $x_{.,\Tindoor}$, $x_{.,\Toutside}$, $x_{.,\RH}$, and $x_{.,\log(\COtwo)}$, and the intercept $\alpha_{.,DS}$ is selected depending on the day segment being analyzed (Fig.~\ref{fig:windownum} and Section~\ref{sec:data_processing} for details). In this study, we decided to omit the interaction between variables to focus on the simplicity and ease of interpretation of the results.

\begin{figure}[t]\centering
\subfloat[]{\includegraphics[width=.235\textwidth]{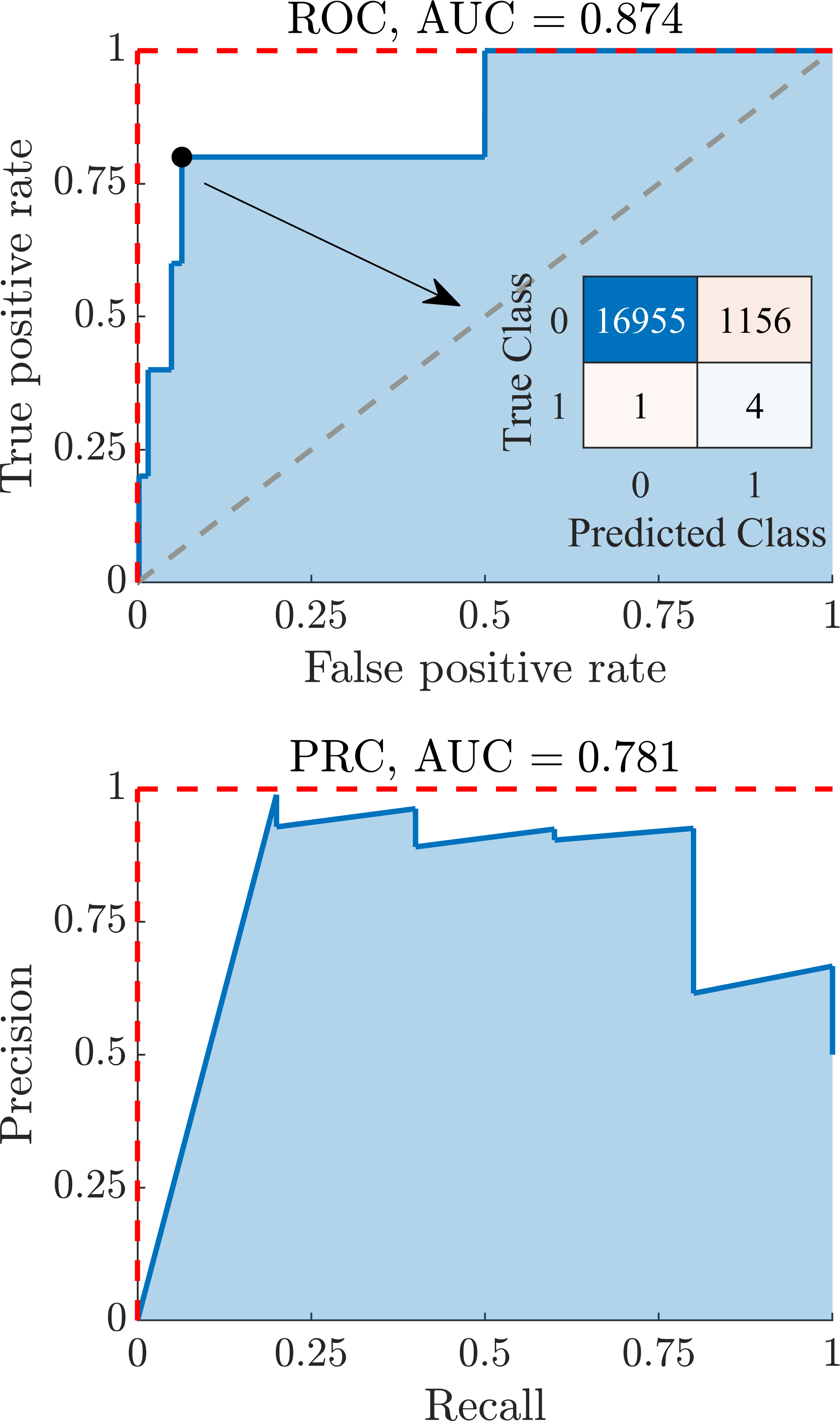}\label{fig:goodness_open}}\;
\subfloat[]{\includegraphics[width=.235\textwidth]{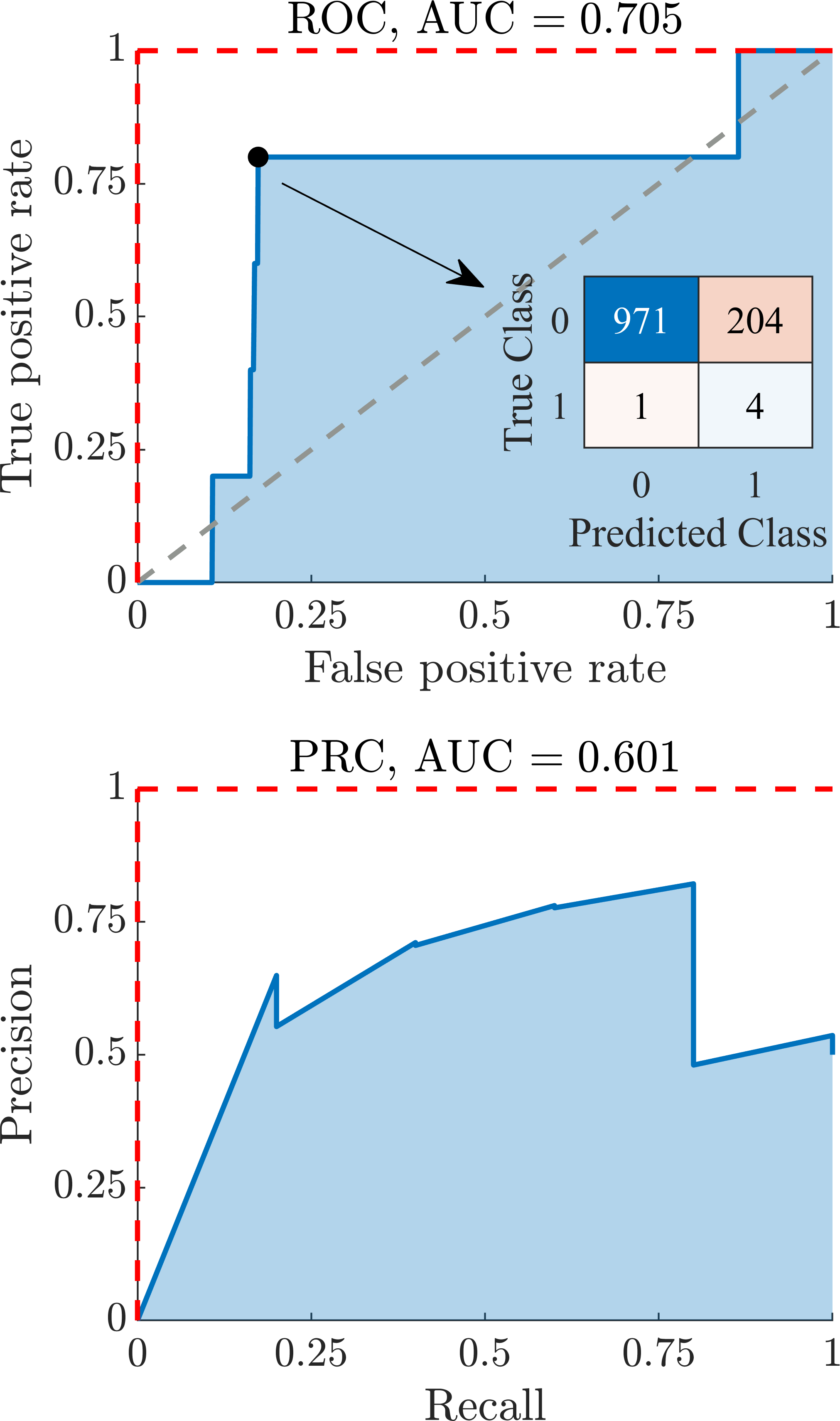}\label{fig:goodness_closed}}
\caption{Goodness of fit for the models of apartment 2 (Section~\ref{sec:results-APT2}). The top panels show the ROC curves and confusion charts (inset panel), presented for the optimal operating point indicated by the black symbol. The bottom panels show the precision-recall curves. Red dashed lines indicate the perfect classifier, and gray dashed lines the baseline classifier. (a): Opening-window model. (b): Closing-window model.}
\label{fig:ROC-PRC}
\end{figure}

\begin{figure}[t]\centering
    \includegraphics[width=0.48\textwidth]{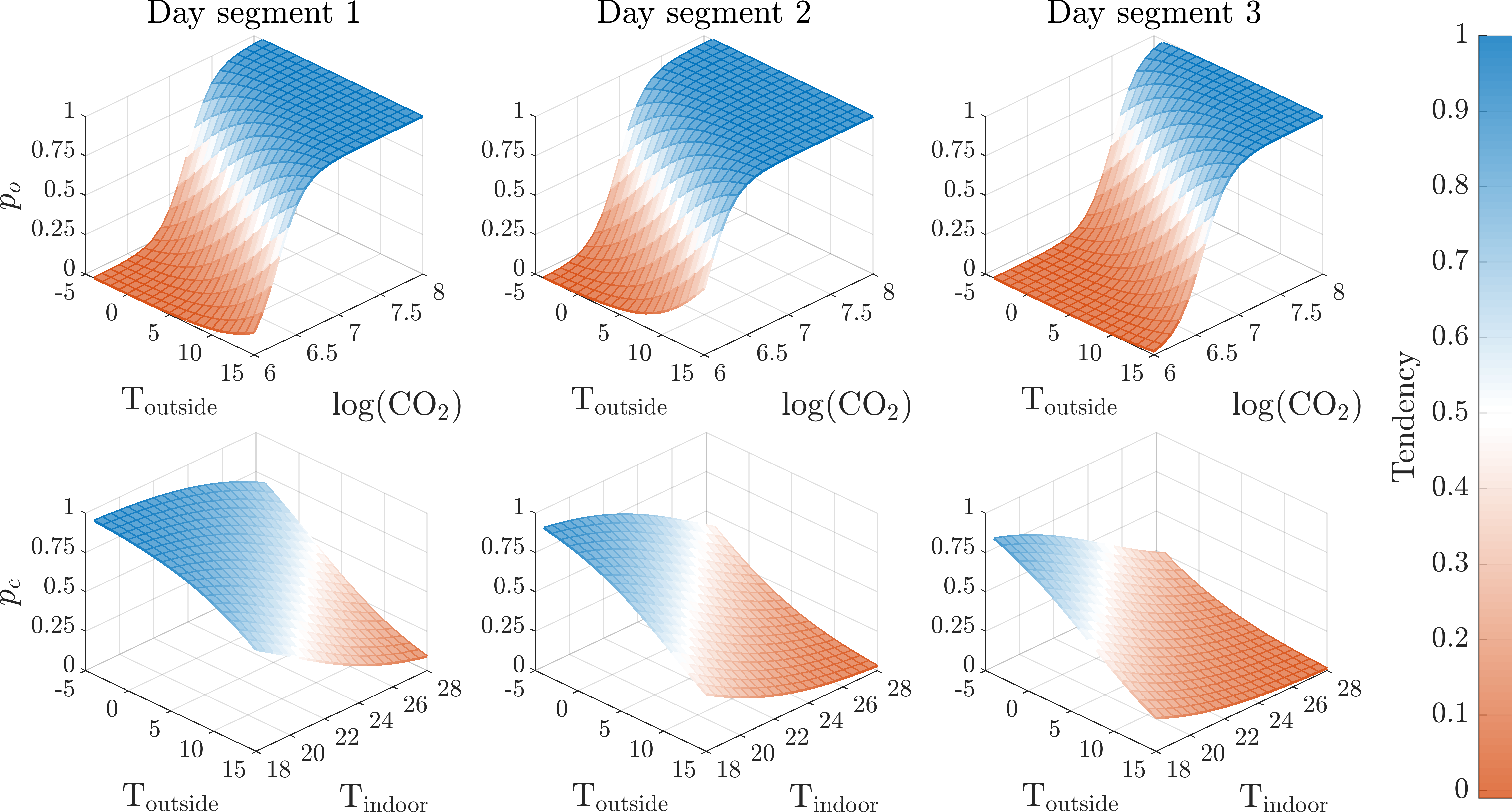}
    \caption{Probability of window-opening action w.r.t. outside temperature and $\log(\COtwo)$ (top panels), and probability of window-closing action w.r.t. indoor and outside temperatures (bottom panels), during different day segments, i.e., morning (DS1), afternoon (DS2), and night (DS3). The surfaces are color-coded based on the tendency of the occupant for apartment 2 towards window-opening or window-closing actions.}
    \label{fig:probability_r2}
\end{figure}

The values of intercept and coefficients of eq.~\eqref{eq:open-win} and eq.~\eqref{eq:close-win} are listed in Table \ref{table:coeffapt2}. 
The data was standardized before employing logistic regression to enable a comparison between the explanatory variables and draw conclusions on their influence on the occupant's behavior. 
 A positive (resp. negative) coefficient indicates a positive (resp. negative) correlation between the associated driver and the opening or closing window action.
Figure~\ref{fig:ROC-PRC} illustrates the goodness of fit of the models in eqs.~\eqref{eq:open-win} and \eqref{eq:close-win}. 
The main key driver characterizing the behavior of the resident of apartment 2 is the time of the day, implying unlikeliness to open the windows during the night segment (DS3, top panel of Fig.~\ref{fig:probability_r2}), and indicating a preference to close the windows in the morning (DS1, bottom panel of Fig.~\ref{fig:probability_r2}). 

The developed model for window-opening action demonstrates a high sensitivity of the occupant to $\COtwo$ (poor air quality drives the occupant towards window-opening behavior) and susceptibility to $\Toutside$ (a warmer outside temperature drives the occupant towards window-opening action and hinders towards window-closing action). The negative correlation between indoor temperature and window-opening action could be explained by the occupant's attitude of keeping the indoor temperature roughly constant and below a potential discomfort threshold. Indeed, the range $\Tindoor$ belongs to in apartment 2 is lower compared to other occupants in KTH Live-In Lab (Fig.~\ref{fig:box-plot}).

\begin{table}[t]\setlength{\tabcolsep}{5.5pt}
\caption{Estimated intercepts and coefficients of the window-opening (eq.~\eqref{eq:open-win}) and window-closing (eq.~\eqref{eq:close-win}) models of apartment 2.}
\centering
\begin{tabular}{ccccccc}\toprule
 & \multicolumn{3}{c}{\textbf{Opening action}} & \multicolumn{3}{c}{\textbf{Closing action}} \\[-1pt]
\cmidrule(rl){2-4} \cmidrule(rl){5-7}
Drivers & Coeff.
& \multicolumn{2}{c}{Conf. interval}& Coeff.
& \multicolumn{2}{c}{Conf. interval}\\
 &  & 2.5\% & 97.5\%&  & 2.5\% & 97.5\%\\[-2pt]\midrule
$\alpha_\mathrm{DS1}$ & -1.729 &  -1.776  & -1.682
&  1.184 & 0.936 & 1.432\\
$\alpha_\mathrm{DS2}$ & -0.242 & -0.271  & -0.212 & 0.023 & -0.099 & 0.143\\
$\alpha_\mathrm{DS3}$ & -3.720 & -3.793  & -3.646& -0.728 & -0.873 & -0.583\\
$\beta_{\Tindoor}$ & -0.627 &  -0.664 & -0.591 & -0.244  &-0.305 &  -0.184\\
$\beta_{\log(\COtwo)}$ & 1.095 &  1.073 &1.117& - & -& -\\
$\beta_\RH$ & 0.299 & 0.249 & 0.345& 0.564 &  0.4394 & 0.690\\
$\beta_{\Toutside}$ & 0.980 &  0.926 &1.030
 & -0.554 & -0.686 & -0.424\\\bottomrule
\end{tabular}\label{table:coeffapt2}
\vspace{-.2cm}\end{table}

\subsection{Window-opening and window-closing models of all apartments in the KTH Live-In Lab} \label{sec:mainresults}
This section provides an overview of the drivers influencing the behavioral patterns of the occupants at the KTH Live-In Lab, focusing in particular on window-opening actions as clearer trends can be observed across the models. Our findings support the results found in \cite{cali2016analysis,andersen2013window}; for readability, we first discuss the results corresponding to the window-opening models and then to the window-closing models.

\begin{figure}\centering
    \includegraphics[width=0.48\textwidth]{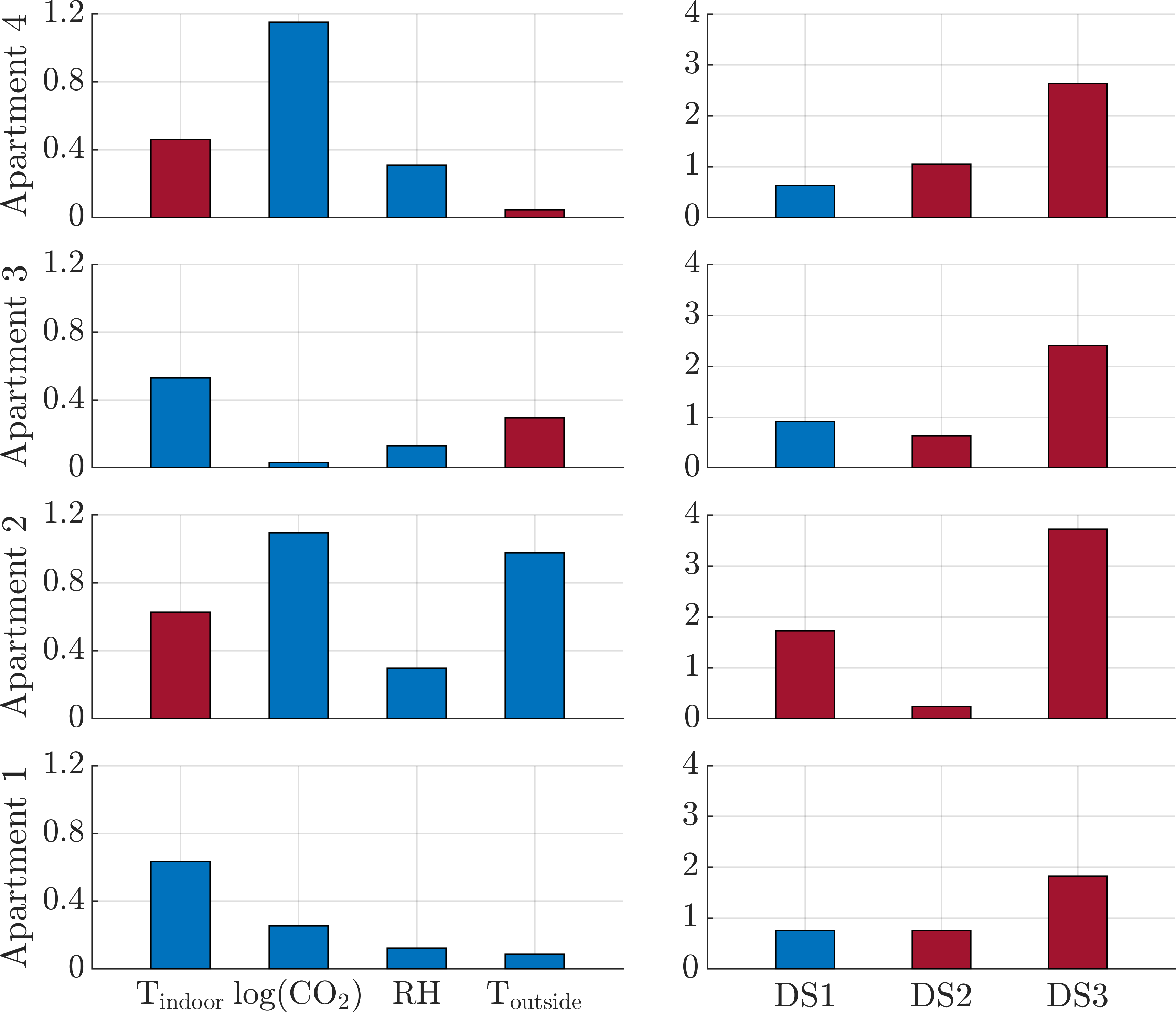}
    \caption{Coefficients and intercept of drivers for window-opening action of all apartments. The blue (resp. red) color indicates a positive (resp. negative) correlation with the window-opening action.}
    \label{fig:AllCoeff_op}
\end{figure}

The magnitude of the effect of physical environmental drivers on occupants' window-opening pattern varies in different apartments and can be interpreted in terms of different habits and personal preferences. For instance, occupants of apartment 1 and apartment 3 seem to be more conscious of the indoor temperature while the occupants of apartment 2 and apartment 4 appear to be more sensitive to the $\COtwo$ concentration levels. However, all the developed models suggest that the occupants are sensitive to air quality (i.e., $\COtwo$ and $\RH$). Further, none of the occupants seems prone to open windows during the night, preferring instead (in general) to open windows in the morning (Fig.~\ref{fig:AllCoeff_op}).

Similar to the window-opening case, occupants' window-closing behavior varies w.r.t. different physical environmental stimuli, while a common factor influencing all residents' behavior is the time of the day. Further, the occupants of apartments 2 and 3 seem both sensitive to indoor and outside temperatures; in particular, the lower the indoor and outside temperatures, the higher the tendency towards a window-closing action. 
However, for the apartments 1 and 4, the AUC of the ROC curve and precision-recall curve are lower than those obtained in the window-opening case, indicating that for these apartments the models are less intuitive, and no significant driver is shown apart from the time of the day.

\begin{remark}
A practical challenge is that our data is highly imbalanced, primarily due to the data period under consideration (winter months), which makes the analysis more complex. Although we considered a 10-fold cross-validation procedure and a balanced weight approach to improve the performance of the logistic regression method, we need to be cautious when evaluating the reliability of the obtained models, even if we obtain high values of AUC of ROC and precision-recall curves (e.g., over 60\% for the models in eqs.~\eqref{eq:open-win} and \eqref{eq:close-win} for apartment 2). In addition, a possible cause for the encountered modeling challenges is the specific layout of the KTH Live-In Lab considered, where the presence of a common area affects the occupancy of apartments.
\end{remark}

\section{CONCLUSION}\label{sec:conclusion}
In this work, we focused on analyzing occupants' behavior patterns related to window operation using real data from a Swedish residential facility, the KTH Live-In Lab. The influence of physical environmental (i.e., indoor and outside temperatures, air quality) and categorical (i.e., time of the day) drivers is captured particularly in the study of the window-opening actions. The findings suggest that, each individual occupant reacts differently to environmental stimuli, while all of the occupants are sensitive to the air quality. Data analysis also indicates a strong correlation between the time of the day and the occupants' tendency to interact with windows (probably related to daily activities or habits). 

This study excluded certain environmental drivers, such as seasonality and solar irradiation, as well as, physiological or psychological drivers, which we plan to take into account in future works, in combination with an extended data collection period. Additionally, future work includes using more advanced statistical methods to deal with heavily imbalanced datasets, as well as analyzing the  occupants' behaviors with other  techniques (e.g., bayesian methods, and conformal prediction methods).

The analysis in this paper is based on data collected during the winter months when window-opening actions highly affect energy consumption. Hence, this work will serve as the foundation to include the dynamical behavior of occupants in designing a data-driven predictive controller \cite{farjadnia2022robust} for HVAC systems in smart buildings to improve energy efficiency while maintaining individual occupants' thermal comfort.

\bibliographystyle{IEEEtran}
\bibliography{setup,addrefs} 

\end{document}